\documentstyle[12pt]{article}

\begin{document}
\parskip 1ex
\setcounter{page}{1}
\oddsidemargin 0pt
 %   Note that \oddsidemargin =  \evensidemargin
\evensidemargin 0pt
\topmargin -40pt
 %    Nominal distance from top of page to  top of \jot = .5ex
%
%%%%%%%%%%%%%%%%%%%%%%%%%%%%%%%%
\newcommand{\be}{\begin{equation}}
\newcommand{\ee}{\end{equation}}
\newcommand{\beq}{\begin{eqnarray}}
\newcommand{\eeq}{\end{eqnarray}}
\def\a{\alpha}
\def\b{\beta}
\def\g{\gamma}
\def\G{\Gamma}
\def\d{\delta}
\def\e{\epsilon}
\def\z{\zeta}
\def\h{\eta}
\def\th{\theta}
\def\k{\kappa}
\def\l{\lambda}
\def\L{\Lambda}
\def\m{\mu}
\def\n{\nu}
\def\x{\xi}
\def\X{\Xi}
\def\p{\pi}
\def\P{\Pi}
\def\r{\rho}
\def\s{\sigma}
\def\S{\Sigma}
\def\t{\tau}
\def\f{\phi}
\def\F{\Phi}
\def\c{\chi}
\def\w{\omega}
\def\W{\Omega}
\def\de{\partial}

%%%%%%%%%%%%%%%%%%%%%%%%%%%%%%%%%%%%
%Without pictures use this macro
\def\pct#1{(see Fig. #1.)}
%%%%%%%%%%%%%%%%%%%%%%%%%%%%%%%%%%%
%With pictures use this macro
%\def\pct#1{\input epsf \centerline{ \epsfbox{#1.eps}}}

%%%%%%%%%%%%%%%%%%%%%%%% FRONT PAGE %%%%%%%%%%%%%%%%%%%%%%%%%%%%%%%%%%%%%
\begin{titlepage}
\hbox{\hskip 12cm ROM2F-97/54  \hfil}
\hbox{\hskip 12cm hep-th/9712176 \hfil}
\hbox{\hskip 12 cm \today}
%\end{flushright}
\vskip 1.8cm
\begin{center} 
{\Large  \bf  Tensor  \ Multiplets \vskip .6cm
in \ Six-Dimensional \ $(2,0)$ \ Supergravity}

\vspace{1.8cm}
 
{\large \large Fabio Riccioni\footnote{I.N.F.N. fellow}}
\vspace{1.2cm}

{\sl Dipartimento di Fisica, \ \
Universit{\`a} di Roma \ ``Tor Vergata'' \\
I.N.F.N.\ - \ Sezione di Roma \ ``Tor Vergata'', \\
Via della Ricerca Scientifica , 1 \ \ \
00133 \ Roma \ \ ITALY}
\end{center}
\vskip 1.8cm

\abstract{We construct the complete coupling of $(2,0)$ supergravity
in six dimensions to $n$ tensor multiplets, extending previous
results to all orders in the fermi fields. The truncation to $(1,0)$
supergravity coupled to tensor multiplets exactly reproduces the complete
couplings recently obtained.}
\vskip 36pt
\hbox{\hskip 1.2cm December  1997 \hfil}
\vfill
\end{titlepage}
%%%%%%%%%%%%%%%%%%%%%%%%%%%%%%%%%%%%%%%%%%%%%%%%%%%%
\makeatletter
\@addtoreset{equation}{section}
\makeatother
\renewcommand{\theequation}{\thesection.\arabic{equation}}
\addtolength{\baselineskip}{0.3\baselineskip} 
%%%%%%%%%%%%%%%%%%%%%%%%%%%%%%%%%%%%%%%%%%%%%%%%%%%%

\section{Introduction}

The massless representations of $(2,0)$ supersymmetry in six dimensions, 
labeled by their
$SU(2)\times SU(2)$ representations, are the gravity multiplet $\bigl(
(1,1)+4(1,\frac{1}{2})+5(1,0) \bigr)$ and the tensor multiplet $\bigl(
(0,1)+4(0,\frac{1}{2})+5(0,0) \bigr)$. In particular $(2,0)$ supergravity
coupled to 21 tensor multiplets is free of gravitational anomalies \cite{agw}, 
and the corresponding six-dimensional vacua naturally arise as 
compactifications of Type-IIB superstrings on $K3$. 

In this paper we construct  the coupling of $(2,0)$
supergravity to an arbitrary number of tensor multiplets, thus
completing \cite{romans} to all orders in the fermi fields.
In \cite{frs} (see also \cite{rs} for a brief review) the 
complete $(1,0)$ supergravity coupled to
tensor and vector multiplets was obtained, generalizing the results of
\cite{ns1} to an arbitrary number of tensors and completing the results
of \cite{as,fms,ns2}. This theory contains reducible gauge
and supersymmetry anomalies induced by tensor couplings, that are 
completely determined solving Wess-Zumino consistency conditions \cite{wz}.
Here we follow the construction of \cite{frs}, but the derivation is simpler 
since in the $(2,0)$ case there are no vector multiplets and 
the low-energy theory does not present a supersymmetry
anomaly. 

In \cite{romans} $(1,0)$ supergravity coupled to $n$ tensor multiplets
was obtained as a truncation of the corresponding $(2,0)$ model to
lowest order in the fermi fields. Here we show how the higher order
fermion terms of the complete $(1,0)$ model \cite{frs} can also be recovered
from a truncation of the complete $(2,0)$ supergravity.
The flat-space limit of $(1,0)$ supergravity coupled to vector and
tensor multiplets was recently studied \cite{dllp}, and two different
global limits were found, but 
in the case of $(2,0)$ supergravity coupled to tensor multiplets the
flat-space limit does not display similar subtleties.

In Section 2 we derive the complete model, and in Section 3 we truncate 
it to $(1,0)$ supergravity. The Appendix explains the notations
and collects some useful identities. 
The conventions follow those used in \cite{frs}.

\section[Complete $(2,0)$ Supergravity in Six Dimensions Coupled to
$n$ Tensor Multiplets]{Complete $(2,0)$ Supergravity in Six Dimensions  \\
Coupled to $n$ Tensor Multiplets}

Chiral extended supersymmetry in six dimensions is generated by four
left spinorial charges $Q^a$ $(a=1,2,3,4)$, obeying the symplectic
Majorana condition
\be
Q^a =\W^{ab} C \bar{Q}^T_b \quad ,
\ee
where $\W^{ab}$ is the invariant tensor of $Sp(4)$. Using the 
isomorphism between $Sp(4)$ and $SO(5)$ the index $a$ can be considered as
an index in the spinorial representation of $SO(5)$.
All fermi fields appear in this representation, and we will denote with $\G^i$ the 
gamma matrices of $SO(5)$.

$(2,0)$ supergravity coupled to $n$ tensor multiplets is described by the 
vielbein $e_\m{}^a$ (now $a$ is a Lorentz index), 
a left-handed gravitino $\psi_\m$, $(n+5)$ antisymmetric 
tensors $B^r_{\m\n}$ $(r=0,...,n+4)$ obeying (anti)self-duality conditions,
$n$ right-handed ``tensorini'' $\chi^m$ $(m=1,...,n)$ and $5n$ scalars.
The scalars parametrize the coset space $SO(5,n)/SO(5) \times SO(n)$,
and are associated to the $SO(5,n)$ matrix
\be
V= \pmatrix{ v^i_r \cr x^m_r } \quad ,
\ee
where $i=1,...,5$ is an index in the vector representation of $SO(5)$.
The matrix elements satisfy the constraints
\beq
& & v^{ir} v^j_r = \delta^{ij} \quad ,\nonumber\\
& & x^{mr} x^n_r = -\delta^{mn} \quad ,\nonumber\\
& & v^{ir} x^m_r =0 \quad, \nonumber\\
& & v^i_r v^i_s - x^m_r x^m_s = \eta_{rs} \quad .\label{scalars}
\eeq
The $SO(5)$ connection is
\be
Q_\m^{ij} =v^i_r (\de_\m v^{j r} ) \quad ,
\ee
while the $SO(n)$ connection is
\be
S_\m^{mn} = (\de_\m x^m_r ) x^{nr} \quad .
\ee

We start considering the theory to lowest order
in the fermi fields \cite{romans}. 
The (anti)self-duality conditions for the tensor fields are \cite{fms}
\be
G_{rs} H^s_{\m\n\r} =\frac{1}{6e} \e_{\m\n\r\a\b\g}H_r^{\a\b\g}
\quad ,
\label{selfdual}
\ee
where 
\be
G_{rs}=v^i_r v^i_s + x^m_r x^m_s \quad .
\ee
These conditions imply that $v^i_r H^r_{\m\n\r}$ are self dual, while 
$x^m_r H^r_{\m\n\r}$ are antiself dual.
The divergence of eq. (\ref{selfdual}) yields the second-order tensor 
equation
\be
D_\m (G_{rs} H^{s \m\n\r} )=0 \label{tensoreq}\quad .
\ee
In \cite{pst} it was shown how to obtain a lagrangian formulation of 
(anti)self-dual two-forms. It involves an additional scalar field and additional
gauge-invariance, and the (anti)self-duality condition,
as well as the elimination of the scalar from the  spectrum, results in a gauge-fixed
version of the complete model. Thus, the lagrangian
\be
e^{-1}{\cal L}_H =\frac{1}{12} G_{rs}H^r_{\m\n\r}H^{s \m\n\r} \quad ,
\ee
that would not make sense for (anti)self-dual forms, is the gauge-fixed version
of the complete tensor lagrangian, and it gives the second-order tensor equation for 
(anti)self-dual two-forms, {\it i.e.} the divergence of the self-duality condition.
With this in mind, we write the lagrangian (see the Appendix for the 
notations)
\beq
& & e^{-1} {\cal L} =-\frac{1}{4} R +\frac{1}{12}G_{rs}H^r_{\m\n\r}H^{s \m\n\r}
+\frac{1}{4}x^{mr}x^{ms}(\de_\m v^i_r ) (\de^\m v^i_s )
-\frac{i}{2}(\bar{\psi}_\m \g^{\m\n\r} D_\n \psi_\r )\nonumber\\
& & \quad \quad -\frac{i}{2}
v^i_r H^{r \m\n\r} ( \bar{\psi}_\m \g_\n \psi_\r )^i +\frac{i}{2} (\bar{\chi}^m \g^\m 
D_\m \chi^m ) -\frac{i}{24}v^i_r H^{r \m\n\r} (\bar{\chi}^m \g_{\m\n\r} \chi^m )^i
\nonumber\\
& & \quad \quad +\frac{1}{2}x^m_r H^{r \m\n\r} (\bar{\psi}_\m \g_{\n\r} \chi^m )
-\frac{1}{2} x^m_r (\de_\n v^{ir} )(\bar{\psi}_\m \g^\n \g^\m \chi^m )^i\quad ,
\label{lag}
\eeq
invariant under the supersymmetry transformations
\beq 
& & \delta e_\m{}^a = -i ( \bar{\e} \g^a \psi_\m ) \quad , \nonumber\\ 
& & \delta B^r_{\m\n} =i v^{ir} ( \bar{\psi}_{[\m} \g_{\n]} \e )^i
+\frac{1}{2} x^{mr} ( \bar{\chi}^m \g_{\m\n} \e ) \quad , \nonumber\\ 
& & \delta v^i_r = x^m_r ( \bar{\e} \chi^m )^i \quad ,\nonumber\\
& & \delta x^m_r = v^i_r (\bar{\e} \chi^m )^i \quad , \nonumber\\
& & \delta \psi_\m = D_\m \e +\frac{1}{4} v^i_r H^r_{\m\n\r} 
\G^i \g^{\n\r} \e \quad , \nonumber\\ 
& & \delta \chi^m =\frac{i}{2} x^m_r
\de_\m v^{ir} \G^i\g^\m \e 
+\frac{i}{12} x^m_r H^r_{\m\n\r} \g^{\m\n\r} \e   
\label{susy}
\eeq 
to lowest order in the fermi fields and using the self-duality condition of eq. 
(\ref{selfdual}). 
The  commutator of two supersymmetry transformations 
on the bosonic fields closes on the local symmetries:
\beq 
& & [ \delta_1 , \delta_2 ] = {\delta}_{gct}( \xi^\m) +
\delta_{tens} (\L^r_\m = -\frac{1}{2} v^{ir} \xi^i_\m -\xi^\n B^r_{\m\n}) 
+\delta_{SO(n)}(A^{mn} = -\xi^\m S_\m^{mn} ) \nonumber \\ 
& & +\delta_{SO(5)} ( A^{ij}=-\xi^\m Q_\m^{ij} )
+\delta_{Lorentz} (\W^{ab} =-\xi_\m \w^{\m a b} +\xi^i_\m v^i_r H^{r \m a b}) \quad ,
\label{susyalg}
\eeq 
where
\be
\xi_\m = -i ({\bar{\e}}_1 \g_\m \e_2 )  \quad ,
\qquad \xi^i_\m = -i ({\bar{\e}}_1 \g_\m \e_2 )^i \quad .
\ee
To this order, one can not  see the local supersymmetry transformation in the gauge
algebra, since the expected parameter,  $\xi^\m \psi_\m$, is generated  by bosonic
variations. As usual, the spin connection satisfies its equation of motion, that to
lowest order in the fermi  fields is 
\be D_\m e_\n{}^a - D_\n e_\m{}^a =0\quad ,
\ee and implies the absence of torsion.

Varying the lagrangian of eq. (\ref{lag}) with respect to the fermi fields one obtains
\be
\g^{\m\n\r} D_\n \psi_\r + v^i_r H^{r \m\n\r} \G^i \g_\n \psi_\r -
\frac{i}{2} x^m_r H^{ r \m\n\r} \g_{\n\r} \chi^m + 
\frac{i}{2} x^m_r \de_\n v^{ir}\G^i \g^\n \g^\m \chi^m =0 
\ee 
and
\be
\g^\m D_\m \chi^m -\frac{1}{12} v^i_r H^{r \m\n\r} \G^i 
\g_{\m\n\r} \chi^m -\frac{i}{2} x^m_r H^{r \m\n\r} \g_{\m\n} \psi_\r-
\frac{i}{2} x^m_r \de_\n v^{ir} \G^i \g^\m \g^\n \psi_\m =0\quad, 
\ee 
while varying it with respect to the scalar fields and to the metric, and considering
only terms without fermions, one obtains
\be 
D_\m (x^m_r \de^\m v^{ir} ) +\frac{2}{3} x^m_r v^i_s 
H^r_{\a\b\g}  H^{s \a\b\g} =0 \quad ,
\ee 
and
\be 
R_{\m\n} -\frac{1}{2} g_{\m\n} R - x^{mr} x^{ms}\de_\m v^i_r \de_\n v^i_s +
\frac{1}{2} g_{\m\n} x^{mr} x^{ms} \de_\a v^i_r \de^\a v^i_s 
- G_{rs} H^r_{\m\a\b} H^s{}_\n{}^{\a\b} =0\quad.
\ee 

Completing these equations will require terms cubic in the fermi fields in the fermionic
equations, and  terms quadratic in the fermi fields in their supersymmetry
transformations.  Supersymmetry will then determine corresponding modifications of the 
bosonic equations, and the (anti)self-duality conditions 
(\ref{selfdual}) will also be
modified by terms quadratic in the fermi fields. Supercovariance actually   fixes  all
terms containing the gravitino in the first-order equations and in the supersymmetry
variations of fermi fields.

The supercovariant forms
\be
\hat{\w}_{\m\n\r} = \w^0_{\m\n\r}  -\frac{i}{2} \lbrace (\bar{\psi}_\m \g_\n \psi_\r )
+(\bar{\psi}_\n \g_\r \psi_\m ) +(\bar{\psi}_\n \g_\m \psi_\r )\rbrace\quad ,
\ee
\beq & & \hat{H}^r_{\m\n\r} = H^r_{\m\n\r}  -\frac{1}{2} x^{mr} \lbrace
( \bar{\chi}^m \g_{\m\n} \psi_\r )+
(\bar{\chi}^m \g_{\n\r}\psi_\m )+( \bar{\chi}^m \g_{\r\m} \psi_\n )\rbrace  \nonumber\\ 
& & \quad \quad - \frac{i}{2} v^{ir}\lbrace
(\bar{\psi}_\m \g_\n \psi_\r )^i+(\bar{\psi}_\n \g_\r
\psi_\m )^i+(\bar{\psi}_\r \g_\m \psi_\n )^i \rbrace \quad ,
\eeq
\be
\hat{\de_\m v^i_r} = \de_\m v^i_r -x^m_r (\bar{\chi}^m \psi_\m )^i \quad ,
\ee 
where
\be
\w^0_{\m\n\r} =\frac{1}{2} e_{\r a}(\de_\m e_\n{}^a -\de_\n e_\m{}^a ) -\frac{1}{2} e_{\m
a}(\de_\n e_\r{}^a -\de_\r e_\n{}^a) +\frac{1}{2} e_{\n a} (\de_\r e_\m{}^a - \de_\m
e_\r{}^a )
\ee 
is the standard spin connection in the absence of  torsion, do not generate
derivatives of the  parameter under supersymmetry. So one can consider
the supercovariant transformations 
\beq 
& & \delta \psi_\m = \hat{D}_\m \e +\frac{1}{4} v^i_r \hat{H}^r_{\m\n\r} \G^i
\g^{\n\r} \e \quad , \nonumber \\ 
& & \delta \chi^m = \frac{i}{2} x^m_r (\hat{\de_\m v^{ir}} ) \G^i \g^\m
\e +\frac{i}{12} x^m_r \hat{H}^r_{\m\n\r} \g^{\m\n\r} \e \quad . \label{newsusy}
\eeq 
The tensorino transformation is complete, while the gravitino  transformation could
include additional terms quadratic in the tensorini. On the other hand, one does not
expect modifications of  the bosonic transformations in the complete theory.

The algebra (\ref{susyalg}) has been obtained varying only the fermi fields in the
bosonic supersymmetry transformations. The next step is to compute  the commutator
completely, varying the bosonic fields as well.  
On $v^i_r$, $x^m_r$ and on the vielbein $e_\m{}^a$ this only modifies  
the local Lorentz parameter, 
\be
\W^{ab}= -\xi^\m \hat{\w}_\m{}^{ab} +\xi^{i\m} v^i_r \hat{H}^r_\m{}^{ab}\quad ,
\ee 
the $SO(n)$ parameter, 
\be
A^{mn} = -\xi^\m S_\m^{mn} +(\bar{\chi}^m \e_2 )^i (\bar{\chi}^n \e_1 )^i -
(\bar{\chi}^m \e_1 )^i (\bar{\chi}^n \e_2 )^i \quad ,
\ee
and the $SO(5)$ parameter, 
\be
A^{ij}= -\xi^\m Q_\m^{ij} +(\bar{\chi}^m \e_2 )^i (\bar{\chi}^m \e_1 )^j -(\bar{\chi}^m
\e_1 )^i (\bar{\chi}^m \e_2 )^j
\ee
together with the inclusion in the algebra of a supersymmetry transformation of 
parameter
\be
\zeta = \xi^\m \psi_\m \quad .
\ee 
These results are obtained using the torsion equation for $\hat{\w}$,
\be
\hat{D}_\m e_\n{}^a - \hat{D}_\n e_\m{}^a = 2S^a{}_{\m\n}= -i(\bar{\psi}_\m
\g^a \psi_\n ) \quad .
\ee

New results come from the complete commutator on $B^r_{\m\n}$, where one needs to use the
(anti)self-duality conditions. 
First of all, since these conditions are first-order equations, they must be 
supercovariant. In general one can require that the tensor
\be
\hat{\cal H}^{r}_{\m\n\r}=
\hat{H}^r_{\m\n\r} +i\a v^r (\bar{\chi}^m \g_{\m\n\r}\chi^m )\quad , \label{newtensor}
\ee 
with $\a$ real coefficient, satisfy the (anti)self-duality conditions 
\be  
G_{rs} \hat{\cal H}^s_{\m\n\r} =\frac{1}{6e} \e_{\m\n\r\a\b\g} \hat{\cal H}_r^{\a\b\g}
\quad .\label{selfdual2}
\ee 
Using eqs.
(\ref{scalars}), one can see that the new $\chi^2$ terms contribute only to the
self-duality condition, while the tensors
$x^m_r \hat{H}^r_{\m\n\r}$ remain antiself dual without extra $\chi^2$  terms.
Consequently, the commutator on
the tensor fields  generates all local symmetries in the proper form, aside from the 
extra terms
\beq 
& & [\delta_1 , \delta_2 ]_{extra} B^r_{\m\n}  =
\frac{\a}{4}v^{ir} [\bar{\e}_2 \g_{\m\r\sigma} \e_1 ]^{ij} (\bar{\chi}^m
\g_\n{}^{\r\sigma} \chi^m )^j -\frac{\a}{4}v^{ir}
[\bar{\e}_2 \g_{\n\r\sigma} \e_1 ]^{ij} (\bar{\chi}^m \g_\m{}^{\r\sigma} \chi^m )^j
\nonumber\\
& & \qquad \qquad \qquad 
+\frac{1}{2}v^{ir} (\bar{\e}_1 \chi^m
)^i (\bar{\chi}^m \g_{\m\n} \e_2 ) -\frac{1}{2} v^{ir} (\bar{\e}_2 \chi^m )^i
(\bar{\chi}^m \g_{\m\n} \e_1 ) \quad ,
\eeq 
that may be canceled adding $\chi^2$ terms to the transformation
of the gravitino. The
most general expression one can add is
\beq
& & \delta^\prime \psi_\m  = ia \ (\bar{\chi}^m \g_{\m\n\r} \chi^m )\g^{\n\r} \e
+ib \ (\bar{\chi}^m \g_{\m\n\r} \chi^m )^i \G^i \g^{\n\r} \e \nonumber\\
& & \qquad +ic \ [\bar{\chi}^m \g_\m
\chi^m ]^{ij} \G^{ij} \e +id \ [\bar{\chi}^m \g^\n \chi^m ]^{ij} \G^{ij} \g_{\m\n} \e
\quad ,
\eeq 
with $a$, $b$, $c$ and $d$ real coefficients, and the total commutator on $B^r_{\m\n}$
then determines all these parameters together with the parameter $\a$ to be
\be 
a=-\frac{1}{64}\quad , \qquad b=-\frac{1}{64} \quad ,\qquad c=\frac{3}{64}\quad ,
\qquad d =-\frac{1}{64} \quad , \qquad \a=-\frac{1}{8} 
\quad .
\ee 
The commutator on $e_\m{}^a$ now closes with a local Lorentz parameter modified by
the addition of
\be
\Delta{\W}^{ab}  =-\frac{i}{16}(\bar{\chi}^m \g^{a b \r}\chi^m )\xi_\r -\frac{i}{16}
(\bar{\chi}^m \g^{a b \r}\chi^m )^i \xi^i_\r +\frac{i}{32}[\bar{\chi}^m \g_\r
\chi^m ]^{ij} \xi^{ij \ ab\r}
\quad , 
\ee 
where
\be
\xi^{ij}_{\m\n\r}=-i[\bar{\e}_1 \g_{\m\n\r} \e_2 ]^{ij} \quad ,
\ee
while the commutators on the scalar fields are not modified.

One can now start to compute the commutators on fermi fields, that as usual close only
on shell. Following \cite{schwarz}, we will actually use this result to  derive the
complete fermionic equations. Let us begin with the  commutator on the tensorini, using
eq. (\ref{newsusy}). Supercovariance determines the field equation of the tensorini up to a term
proportional to
$\chi^3$. Closure of  the algebra fixes this additional term, and the end result is
\beq 
& & \g^\m \hat{D}_\m \chi^m -\frac{1}{12}v^i_r \hat{H}^r_{\m\n\r}\G^i \g^{\m\n\r}
\chi^m -\frac{i}{2} x^m_r \hat{H}^{r \m\n\r} \g_{\m\n} \psi_\r \nonumber \\ &
&-\frac{i}{2} x^m_r ( \hat{\de_\n v^{ir}} )\G^i \g^\m \g^\n \psi_\m  
+\frac{i}{16} [\bar{\chi}^n \g_\m \chi^n ]^{ij} \G^{ij} \g^\m \chi^m = 0
\quad .\label{chieq}
\eeq 
The complete commutator of two supersymmetry transformations on the tensorini is
then
\beq 
& & [\delta_1 ,\delta_2 ] \chi^m = \delta_{gct} \ \chi^m +\delta_{Lorentz} \ \chi^m
+\delta_{SO(n)} \ \chi^m +\delta_{SO(5)} \ \chi^m +\delta_{susy} \ \chi^m\nonumber\\
& & \qquad \qquad + \frac{3}{8} \xi_\a \g^\a  \ [{\rm eq.} \ \chi^m ] -\frac{1}{8}
\xi^i_\a \G^i \g^\a \ [{\rm eq.} \ \chi^m ] \quad .
\eeq 
A similar result can be obtained for the gravitino. In this case the complete
equation,
\beq 
& & \g^{\m\n\r} \hat{D}_\n \psi_\r +\frac{1}{4} v^i_r \hat{H}^r_{\n\a\b}\G^i
\g^{\m\n\r} \g^{\a\b} \psi_\r -\frac{i}{2}x^m_r \hat{H}^{r \m\n\r} \g_{\n\r}
\chi^m  +\frac{i}{2} x^m_r (\hat{\de_\n v^{ir}})\G^i 
\g^\n \g^\m \chi^m \nonumber \\ 
& & +\frac{i}{8}(\bar{\chi}^m\g^{\m\n\r}\chi^m )\g_\n \psi_\r +\frac{i}{32}
(\bar{\chi}^m \g_{\n\r\sigma }\chi^m )\g^{\m\n\r}\psi^\sigma -\frac{i}{32}
(\bar{\chi}^m \g^{\m\n\r}\chi^m ) \g_{\sigma\n\r}\psi^\sigma \nonumber\\ 
& & +\frac{i}{8}(\bar{\chi}^m \g^{\m\n\r}\chi^m )^i \G^i \g_\n \psi_\r 
+\frac{i}{32} (\bar{\chi}^m \g_{\n\r\sigma }\chi^m )^i \G^i \g^{\m\n\r}\psi^\sigma 
-\frac{i}{32} (\bar{\chi}^m \g^{\m\n\r}\chi^m )^i \G^i \g_{\sigma\n\r}\psi^\sigma 
\nonumber\\ 
& & +\frac{i}{16}[\bar{\chi}^m \g_\n \chi^m ]^{ij} \G^{ij} \g^\m \psi^\n
-\frac{i}{16}[\bar{\chi}^m \g^\m \chi^m ]^{ij} \G^{ij} \g^\n \psi_\n =0 \quad ,
\label{gravitinoeq}
\eeq 
is fixed by  supercovariance, and the commutator closes up to  terms proportional
to a particular combination of eq. (\ref{gravitinoeq}) and its $\g$-trace. Precisely,
one obtains
\beq 
& & [\delta_1 ,\delta_2 ] \psi_\m = \delta_{gct} \ \psi_\m
+\delta_{Lorentz} \ \psi_\m +\delta_{susy} \ \psi_\m +\delta_{SO(5)} \ \psi_\m
\nonumber\\ 
& & +\frac{5}{16} \xi_\sigma \g^\sigma \ [eq. \ \psi_\m ]-\frac{7}{64}
\xi_\sigma \g^\s \g_\m \ [\g -trace]+\frac{1}{4}\xi_\sigma \g_\m \lbrace
[eq. \ \psi^\sigma]-\frac{1}{4}\g^\sigma [\g-trace] \rbrace \nonumber\\
& & +\frac{1}{16} \xi^i_\sigma \G^i \g^\sigma \ [eq. \ \psi_\m ]+\frac{1}{64}
\xi^i_\sigma \G^i \g^\s \g_\m \ [\g -trace]-\frac{1}{4}\xi^i_\sigma 
\G^i \g_\m \lbrace
[eq. \ \psi^\sigma]-\frac{1}{4}\g^\sigma [\g-trace] \rbrace \nonumber\\
& & -\frac{1}{384}\xi^{ij}_{\sigma \delta\tau}\G^{ij} \g^{\sigma\delta\tau}
\lbrace [eq. \ \psi_\m]-\frac{1}{4}\g_\m [\g-trace] \rbrace 
\quad .
\label{gravitinoalg}
\eeq 

Summarizing, from the algebra we have obtained the complete fermionic equations of
$(2,0)$ six-dimensional supergravity coupled to $n$ tensor multiplets. In addition, the
modified 3-form
\be
\hat{\cal H}^{r}_{\m\n\r}=
\hat{H}^r_{\m\n\r} -\frac{i}{8}v^{ir} (\bar{\chi}^m \g_{\m\n\r}\chi^m )^i
\ee 
satisfies the (anti)self-duality conditions
\be 
G_{rs} \hat{\cal H}^{s}_{\m\n\r}=\frac{1}{6e}\e_{\m\n\r\a\b\g}
\hat{\cal H}^{\a\b\g}_r \quad. \label{finalselfdual}
\ee
We have also identified the complete supersymmetry transformations, 
that we collect here for convenience:
\beq 
& & \delta e_\m{}^a =-i(\bar{\e} \g^a \psi_\m ) \quad,\nonumber\\ 
& & \delta B^r_{\m\n} =i v^{ir} (\bar{\psi}_{[\m} \g_{\n]} \e )^i+ 
\frac{1}{2} x^{mr} (\bar{\chi}^m \g_{\m\n} \e ) \quad, \nonumber\\ 
& & \delta v^i_r = x^m_r (\bar{\chi}^m \e )^i \quad,\nonumber\\ 
& & \delta x^m_r = v^i_r (\bar{\chi}^m \e )^i \quad ,\nonumber\\
& & \delta \psi_\m =\hat{D}_\m \e +\frac{1}{4} v^i_r \hat{H}^r_{\m\n\r} \G^i
\g^{\n\r}\e -\frac{i}{64}( \bar{\chi}^m \g_{\m\n\r}\chi^m )\g^{\n\r}\e
-\frac{i}{64}( \bar{\chi}^m \g_{\m\n\r}\chi^m )^i \G^i \g^{\n\r}\e \nonumber\\
& & \qquad +\frac{3i}{64}
[\bar{\chi}^m \g_\m \chi^m]^{ij} \G^{ij} \e -\frac{i}{64}[\bar{\chi}^m \g^\n \chi^m 
]^{ij} \G^{ij} \g_{\m\n} \e
\quad ,\nonumber\\ 
& & \delta \chi^m = \frac{i}{2} x^m_r (\hat{\de_\a v^{ir}} )\G^i \g^\a \e +
\frac{i}{12} x^m_r \hat{H}^r_{\a\b\g} \g^{\a\b\g}\e \quad .
\eeq 

From the equations of the fermi fields one obtains the complete Lagrangian 
\beq
& &  e^{-1} {\cal{L}}  = -\frac{1}{4} R +\frac{1}{12}G_{rs}H^r_{\m\n\r}H^{s \m\n\r}
+\frac{1}{4}x^{mr}x^{ms}(\de_\m v^i_r ) (\de^\m v^i_s )\nonumber\\
& & \quad \quad -\frac{i}{2}(\bar{\psi}_\m \g^{\m\n\r} D_\n [\frac{1}{2}(\w
+\hat{\w} )]
\psi_\r ) -\frac{i}{8}v^i_r [H+\hat{H}]^{r \m\n\r}(\bar{\psi}_\m \g_\n \psi_\r)^i
\nonumber \\ 
& & \quad \quad +\frac{i}{48} v^i_r [H+\hat{H} ]^r_{\a\b\g} (\bar{\psi}_\m
\g^{\m\n\a\b\g}\psi_\n )^i+\frac{i}{2} (\bar{\chi}^m \g^\m D_\m (\hat{\w})
\chi^m )\nonumber \\ 
& & \quad\quad -\frac{i}{24}v^i_r \hat{H}^r_{\m\n\r} (\bar{\chi}^m \g^{\m\n\r}
\chi^m )^i +\frac{1}{4}x^m_r [\de_\n v^{ir} +\hat{\de_\n v^{ir}} ](\bar{\psi}_\m \g^\n
\g^\m \chi^m)^i \nonumber\\ 
& & \quad \quad -\frac{1}{8} x^m_r [H+\hat{H}]^{r \m\n\r} ( \bar{\psi}_\m
\g_{\n\r}
\chi^m )+\frac{1}{24}x^m_r [H+\hat{H}]^{r \m\n\r} (\bar{\psi}^\a \g_{\a\m\n\r}
\chi^m ) \nonumber\\ 
& & \quad \quad +\frac{1}{8}(\bar{\chi}^m \g^{\m\n\r} \chi^m )(\bar{\psi}_\m
\g_\n \psi_\r )-\frac{1}{64}[\bar{\chi}^m \g_\m \chi^m ]^{ij}
[\bar{\chi}^n \g^\m \chi^n ]^{ij}\quad ,
\label{fermilag}
\eeq 
where, in the 1.5 order formalism,  the spin connection
\beq & & \w_{\m\n\r} =\w^0_{\m\n\r} -\frac{i}{2}\lbrace (\bar{\psi}_\m \g_\n \psi_\r )
+(\bar{\psi}_\n \g_\r
\psi_\m)+(\bar{\psi}_\n \g_\m \psi_\r ) \rbrace\nonumber\\ 
& & \qquad -\frac{i}{4}(\bar{\psi}^\a 
\g_{\m\n\r\a\b} \psi^\b )-\frac{i}{4} (\bar{\chi}^m \g_{\m\n\r} \chi^m )
\label{1.5}
\eeq 
satisfies its equation of motion, and is thus kept fixed in all variations.

As we anticipated, varying
${\cal{L}}$  with respect to the antisymmetric tensor
$B^r_{\m\n}$ yields  the second-order tensor equation, the divergence of eq.
(\ref{finalselfdual}).

Supersymmetry is finally proved showing that
\be 
\delta F \frac{\delta {\cal{L}}}{\delta F}+\delta B 
\frac{\delta {\cal{L}}}{\delta B}=0
\quad,\label{susyproof}
\ee 
where $F$ and $B$ denote collectively the fermi  and  bose fields 
aside from the
antisymmetric tensors, that are constrained to satisfy eq. (\ref{finalselfdual}). 
We would like to stress that the equations for the fermi
fields defined from the gauge algebra differ from the lagrangian equations
by overall factors that may be simply identified.

In order to study the flat-space limit, one has to rescale the fields
following \cite{dllp} and then let the gravitational constant 
$\kappa$ tend to zero. The end result is 
a free field theory of $n$ antiself-dual tensors, $5n$ scalars and $n$
right-handed spinors. In this limit the supersymmetry transformations become
\beq
& & \delta B^m_{\m\n} =\frac{1}{2} (\bar{\e} \g_{\m\n} \chi^m ) \quad, 
\nonumber\\
& & \delta \phi^{mi} =(\bar{\e} \chi^m )^i \nonumber\\
& & \delta \chi^m =-\frac{i}{2} (\de_\m \phi^{mi} )\G^i \g^\m \e
+\frac{i}{12}H^m_{\m\n\r} \g^{\m\n\r} \e \quad ,
\eeq 
and the algebra closes on the equations of $\chi^m$ to give 
a translation and a gauge transformation.

\section{Truncation to $(1,0)$ Supergravity}

In \cite{romans} $(1,0)$ supergravity coupled to $n$ tensor multiplets 
was obtained to lowest order in the fermi fields as a truncation of $(2,0)$ 
supergravity. 
Here we generalize the construction, showing 
how the truncation acts on the $Sp(4)$ bilinears to give the $Sp(2)$ bilinears of ref. 
\cite{frs}. 

The scalars $v^i_r$, as well as the bilinears $(\bar{\psi}\chi )^i$, where $\psi$
and $\chi$ are generic spinors of different chirality, loose their $Sp(4)$ index
and give $v_r$ and $(\bar{\psi}\chi )$ in the notations of \cite{frs}.
In this way, for example, one can obtain the supersymmetry transformations of 
$\psi_\m$ and $\chi^m$ to lowest order in the fermi fields \cite{romans}.
On the other hand, the product of two bilinears with anomalous behavior under 
Majorana-flip, 
$$
[\bar{\psi}\chi ]^{ij} [\bar{\l} \e ]^{ij} \quad ,
$$ 
becomes 
$$  
-4[\bar{\psi}\chi ]^i [\bar{\l}\e ]^i 
$$ 
in $(1,0)$ notations. 
Following these rules, the complete supersymmetry transformation of the gravitino 
becomes in the $(1,0)$ model
\beq
& & \delta \psi_\m =\hat{D}_\m \e +\frac{1}{4} v_r \hat{H}^r_{\m\n\r} 
\g^{\n\r}\e -\frac{i}{32}( \bar{\chi}^m \g_{\m\n\r}\chi^m )\g^{\n\r}\e
 \nonumber\\
& & \qquad -\frac{3i}{16}
[\bar{\chi}^m \g_\m \chi^m]^{i} \sigma^i \e +\frac{i}{16}[\bar{\chi}^m \g^\n \chi^m 
]^{i} \sigma^i \g_{\m\n} \e
\quad ,
\eeq
that coincides with the fourth of eqs. (2.38) of \cite{frs} after some Fierz 
rearrangement.
In the same way one can show that all other truncations correspond to terms
already found in \cite{frs}. 

\section*{Acknowledgment}

It is a pleasure to thank A. Sagnotti for several stimulating 
discussions and for a careful reading of the manuscript. I am also grateful
to K. Ray for helpful discussions.

\section{Appendix}

Spinors satisfy the symplectic Majorana condition
\be
\psi^a =\W^{ab} C \bar{\psi}^T_b \quad,\label{majoranacond}
\ee where
\be
\bar{\psi}_a =(\psi^a )^\dagger \g_0 \quad .
\ee 
Any bilinear
$\bar{\psi}_a
\chi^b$ carries a pair of $Sp(4)$ indices, and can be decomposed in terms of the
identity, the five $\G^i$ matrices and the ten $\G^{ij}$ matrices. 
Indeed, one can form the bilinears
\be 
(\bar{\psi} \chi )=\bar{\psi}_a \chi^a \quad ,\qquad (\bar{\psi} \chi )^i =
 \G^{ia}{}_b \bar{\psi}_a \chi^b \quad,\qquad [\bar{\psi} \chi ]^{ij}
=\G^{ija}{}_b \bar{\psi}_a \chi^b\quad , 
\ee 
and using the $SO(5)$ Clifford algebra one has
\be
\bar{\psi}_b \chi^a =\frac{1}{4}\delta^a_b (\bar{\psi} \chi ) +\frac{1}{4}
\G^{ia}{}_b [\bar{\psi} \chi ]^i- \frac{1}{8}\G^{ija}{}_b
[\bar{\psi}\chi ]^{ij} \quad.\label{sp4exp}
\ee 
Using eq. (\ref{majoranacond}), one can then see that the fermi bilinears
$(\bar{\psi} \chi )$ and $(\bar{\psi}\chi )^i$ have standard behavior under Majorana
flip, namely
\be 
(\bar{\psi} \chi ) =(\bar{\chi} \psi ) \quad ,\qquad (\bar{\psi}\chi )^i = 
(\bar{\chi}\psi )^i \quad ,
\ee 
while all ten bilinears $[\bar{\psi} \chi ]^{ij} $ have the anomalous behavior
\be 
[\bar{\psi} \chi ]^{ij} = -[\bar{\chi} \psi ]^{ij} \quad .
\ee 

One can study Fierz relations between spinor bilinears  using eq. (\ref{sp4exp}). In
particular,  if $\psi$ and $\chi$ have the same chirality, one has 
\be
\psi^a \bar{\chi}_b -\chi^a \bar{\psi}_b = -\frac{1}{8} (\bar{\chi} \g^\a \psi
)\delta_b^a \g_\a -\frac{1}{8} (\bar{\chi} \g^\a \psi )^i \G^{ia}{}_b \g_\a
-\frac{1}{192}  [\bar{\chi} \g^{\a\b\g} \psi ]^{ij} \G^{ija}{}_b
\g_{\a\b\g} 
\ee
and
\be
\psi^a \bar{\chi}_b +\chi^a \bar{\psi}_b =\frac{1}{16} [\bar{\chi}\g^\a \psi ]^{ij}
\G^{ija}{}_b \g_\a+\frac{1}{96}(\bar{\chi}\g^{\a\b\g} \chi ) \delta^a_b \g_{\a\b\g} 
+\frac{1}{96}(\bar{\chi}\g^{\a\b\g} \chi )^i \G^{ia}{}_b \g_{\a\b\g} \quad .
\ee

\end{document}